\documentclass[aps,prl,twocolumn,showpacs,showkeys,square,numbers,amssymb,amsmath,nobibnotes,footinbib]{revtex4}
\usepackage{docs}%
\usepackage[utf8]{inputenc}%
\usepackage[T1]{fontenc}%
\usepackage{bm}%
\usepackage[colorlinks=true,linkcolor=red]{hyperref}
\usepackage{amsmath}
\usepackage{amsfonts}
\usepackage{amssymb}
\usepackage{verbatim}
\usepackage{epsfig}
\usepackage{eucal}
\usepackage{graphicx}
\usepackage{soul}
\usepackage[sort&compress]{natbib}
\usepackage{times}
\def\g{\gamma}
\begin{document}

%\preprint{}

\title{Entanglement production in non-ideal cavities and optimal opacity}


\author{Dario Villamaina}
\author{Pierpaolo Vivo}

\affiliation{Laboratoire de Physique Th\'{e}orique et Mod\`{e}les
Statistiques (UMR 8626 du CNRS), Universit\'{e} Paris-Sud,
B\^{a}timent 100, 91405 Orsay Cedex (France)}

%\thanks{}


\date{\today}

\begin{abstract}
We compute analytically the distributions of concurrence $\bm{\mathcal{C}}$ and
squared norm $\bm{\mathcal{N}}$ for the production of electronic entanglement in a
chaotic quantum dot. The dot is connected to the external world via
one ideal and one partially transparent lead, characterized by the opacity $\gamma$.
The average concurrence increases with $\gamma$ while the average squared norm of the entangled state decreases, making it less likely to be detected. When a minimal detectable norm $\bm{\mathcal{N}}_0$ is required, the average concurrence is maximal for an optimal value of the opacity $\gamma^\star(\bm{\mathcal{N}}_0)$ which is explicitly computed as a function of $\bm{\mathcal{N}}_0$. If $\bm{\mathcal{N}}_0$ is larger than the critical value $\bm{\mathcal{N}}_0^\star\simeq 0.3693\dots$, the average entanglement production is maximal for the completely ideal case, a direct consequence of an interesting bifurcation effect.
\end{abstract}


\maketitle


%\section{Introduction}

The phenomenon of entanglement between spatially separated particles
is probably one of the most baffling predictions of quantum mechanics.
Initially, the idea that entangled quantum states could be subject to
non-classical correlations was seen as a possible evidence of the
incompleteness of quantum theory \cite{EPR35}.  Nowadays, thanks to
fascinating experimental developments, the reality of quantum
entanglement is no longer disputed. Besides its fundamental and
intellectually challenging interest, modern research on entanglement
mainly focuses on its possible applications to quantum information
processing, cryptography and teleportation \cite{NC-book,alber}. As a
consequence, it is of paramount interest to discover feasible pathways
for the production, manipulation, and detection of quantum
entanglement in a variety of physical devices.

\begin{figure}[htb]
\centerline{\epsfig{ figure=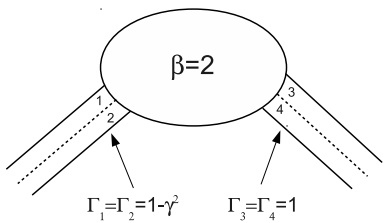 ,width=0.28\textwidth}} \caption{Sketch of the orbital entangler. \label{figastronave}}
\end{figure}

Among the most promising routes, the entanglement in solid-state
electronic systems \cite{B06,B07} plays a prominent role.  As far as
the production of entangled electrons is concerned, several proposals
based on interacting \cite{inter} and noninteracting \cite{noninter,
  FMF06,BKMY04} electron mechanisms are already available.  A very
interesting interactionless mechanism was recently proposed by
Beenakker \textit{et al.} \cite{BKMY04} where a ballistic quantum dot
(see sketch in Fig.~\ref{figastronave}) is used as an orbital
entangler for pairs of noninteracting electrons \cite{samuelsson}. A
quantum dot is basically a mesoscopic electronic billiard connected to
the external world by two double-channel leads and brought out of
equilibrium by an applied external voltage. Assuming that the
classical electron dynamics inside the cavity is chaotic, experimental
observables such as conductance and shot noise fluctuate from sample
to sample and their statistics (about which nearly everything is known
\cite{cond}) is governed by the transmission eigenvalues of the
cavity.

In order to quantify the degree of entanglement produced inside the
cavity, one uses a standard measure $\bm{\mathcal{C}}$ called
\emph{concurrence}, which is related to the violation of a Bell
inequality for current correlators. The average and variance of the
concurrence for a chaotic dot \emph{with ideal leads} (i.e. when the
tunnelling probability across the leads is unity) was computed in
\cite{BKMY04}. It was found that these two moments are practically
unaffected by the breaking of time-reversal invariance (TRI), but the
fluctuations can be very large (of the same order as the
average). This implies that a statistical description of entanglement
production based on just the first moments is highly inaccurate. The
full distribution of the concurrence (again in the case of ideal
leads) was computed by Gopar and Frustaglia \cite{gopar}, and indeed
they found remarkable differences in the distributions between the
cases with preserved $(\beta=1)$ and broken $(\beta=2)$ TRI (here
$\beta$ denotes the Dyson index of the cavity), although the first
moments are only very mildly affected. Still in the domain of ideal
cavities, geometrical constraints on the entanglement production were
discovered in \cite{novaes}. The explicit calculation of the joint
distribution of concurrence and the squared norm $\bm{\mathcal{N}}$ of
the entangled state yields the relation
$\bm{\mathcal{N}}(1+\bm{\mathcal{C}})<1$, which implies that more
entangled states are less likely to be detected as they are bound to have a smaller norm. This conclusion
holds irrespective of whether the leads are ideal or not. However, in the case of non-ideal leads considered here, the above inequality brings intriguing consequences that will be discussed below.

We remark at this stage that all available analytical results to date heavily rely on the assumption that
all leads are ideal. This restriction is however hardly tolerable, as
ideal transparencies are never realized in experiments.  The
tunneling probabilities ($\Gamma_\nu$, $\nu=1,2$) between the leads
and the cavity can be tuned, and it is therefore of great interest to study the entanglement production as a function of
$\Gamma_\nu$. Besides the experimental relevance \cite{butt},
this investigation is fascinating from a purely theoretical point of
view since numerical simulations suggest that uneven contact transparencies may
confer advantages in the production of entangled states \cite{brasil} (see also \cite{almeidamacedo} for a numerical study of the distribution of charge cumulants in non-ideal cases).  

The purpose of this Letter is to compute analytically the
joint distribution $P_\gamma(\bm{\mathcal{C}},\bm{\mathcal{N}})$ of
concurrence and squared norm of the entangled state for $\beta=2$ (broken TRI) in cavities
supporting one ideal and one non-ideal lead characterized by the
\emph{opacity} $\gamma\in [0,1]$ (related to the tunneling probability via $\gamma=\sqrt{1-\Gamma}$). All the known results in the ideal setting are easily recovered as a limiting $\gamma\to 0$ case of our theory, whose predictions might be possibly tested within current experimental capabilities. We find that the non-ideality of the opaque lead is responsible for a rich and interesting behavior of the entanglement production process, which can be optimized by carefully tuning $\g$ on a critical value $\g^\star$ that we explicitly compute. This is due to a nontrivial interplay between an \emph{increase} in the average concurrence and a simultaneous \emph{decrease} in the average squared norm of the entangled state as the opacity $\g$ is pulled away from ideality. This calculation is made possible thanks to a recent breakthrough by Vidal and Kanzieper \cite{vidal}, who were
able to compute the joint distribution of transmission eigenvalues for the case of a
$\beta=2$ cavity with $n_L$ non-ideal channels in the left lead and
$n_R$ ideal channels in the right lead (see \cite{supp} for details). Before summarizing our results, let us first describe in detail the experimental setting and the relevant theoretical framework.

The orbital entangler we consider, first proposed in Ref.~\cite{BKMY04} is
sketched in Fig.~\ref{figastronave}.  It consists of a quantum dot
with two attached double-channel leads at the left and right. Each
lead is connected to an electron reservoir. Applying a small voltage
between the two reservoirs gives rise to a electronic current flowing
through the dot from left to right. The scattering within the dot
leads to the production of entangled pairs between transmitted (to the
right) and reflected (to the left) electrons.  More precisely, the
outgoing state $|\psi_{\mathrm{out}}\rangle$ of two scattered
electrons can be written as the superposition
$|\psi_{\mathrm{out}}\rangle =|\psi_{\ell\ell}\rangle
+|\psi_{rr}\rangle +|\psi_{\ell r}\rangle$, where
$|\psi_{\ell\ell}\rangle$ and $|\psi_{rr}\rangle$ are separable
states, corresponding to both electrons being scattered to the left or
to the right, while $|\psi_{\ell r}\rangle$ (representing a state
where one electron is scattered to the left and the other to the
right) may be nonseparable. Since an electron leaving the quantum dot
to the left side can choose between channels $1$ and $2$ for escaping, this defines a two-level quantum system or
\emph{qubit}. The same happens with an electron escaping to the right
side through leads $3$ and $4$.  This means that the state
$|\psi_{\ell r}\rangle$ in general describes (up to a normalization
factor) a two-qubit entangled state.

A widely used measure for quantifying two-qubit entanglement is the
concurrence $\bm{\mathcal{C}}$ \cite{W98}, an entanglement monotone
constructed from the two-qubit density matrix. The
concurrence is intimately connected to the transport properties of the cavity, and the crucial question is then how to compute it in the proposed setting.

Consider the $4\times 4$ scattering matrix $\bm{\mathcal{S}}$ of the cavity,
\begin{equation}
\bm{\mathcal{S}}=\left[
\begin{array}{cc}
\mathbf{r} & \mathbf{t}'   \\
\mathbf{t} & \mathbf{r}'
\end{array}
\right]\; ,
\label{S}
\end{equation}
where $\mathbf{r},\mathbf{r}',\mathbf{t},$ and $\mathbf{t}'$ are $2
\times 2$ reflection and transmission matrices, respectively. In the
presence of TRI, $\bm{\mathcal{S}}$ is unitary and symmetric. If TRI
is broken (due to, e.g., the application of a magnetic flux), then
$\bm{\mathcal{S}}$ is only unitary. Let $\tau_1$ and $\tau_2$ be the
eigenvalues ($0\leq \tau_i\leq 1$) of the Hermitian matrix
$\mathbf{tt}^\dagger$.  Then, the concurrence $\bm{\mathcal{C}}$ can
be written in terms of the transmission eigenvalues $\tau_1$ and
$\tau_2$ as \cite{BKMY04,B06}
\begin{equation}
\label{concu_t1t2}
\bm{\mathcal{ C}}=\frac{2\sqrt{\tau_1(1-\tau_1)\tau_2(1-\tau_2)}}{\tau_1+\tau_2-2\tau_1\tau_2} .
\end{equation}
We note that the concurrence varies
from 0 to 1. 
 The case ${\bm{\mathcal{C}}}=0$ corresponds to separable
nonentangled states, while maximally entangled (Bell) states
correspond to $\bm{\mathcal{C}}=1$. Those states with a $0 <
\bm{\mathcal{C}} < 1$ are non-separable partly entangled states.
From Eq.~(\ref{concu_t1t2}),
the entanglement is maximal ($\bm{\mathcal{C}}=1$) when $\tau_1 =\tau_2$, 
and minimal ($\bm{\mathcal{C}}=0$) when $\tau_1=0$ and $\tau_2=1$ or $\tau_1=1$ and $\tau_2=0$. A finite value
of $\bm{\mathcal{C}}$ guarantees that the left and right outgoing
channels are orbitally entangled. The denominator of \eqref{concu_t1t2} is the squared norm
\begin{equation}
\bm{\mathcal{N}}=\tau_1+\tau_2-2\tau_1\tau_2
\end{equation}

\begin{figure}[htb]
\begin{center}
\includegraphics[scale=0.30,clip=true]{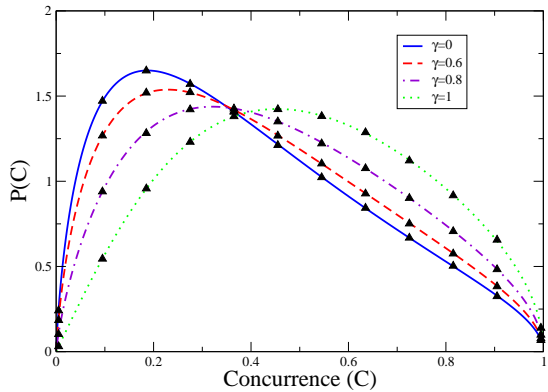}
 \caption{Probability density of the concurrence \eqref{finalC} compared with numerical simulations (black triangles) for different values of $\g$. \label{figprob}}
\end{center}
\end{figure}

The chaotic scattering in the cavity implies that the entanglement
production is basically a stochastic process, governed by a random
scattering matrix \cite{jp1,baranger-mello,jalabert,brouw}. In particular, the concurrence as well as the
squared norm are random variables whose statistics is the central
object of this Letter. From \cite{vidal} we deduce after lengthy algebra \cite{supp} that the joint distribution of the two transmission eigenvalues 
in our setting is given by
\begin{equation}
P_{\gamma,\gamma}(\tau_{1},\tau_{2})=\frac{(\gamma^{2}-1)^{8} \sum_{i,j=0}^{4}A_{ij}(\g)(1-\tau_{1})^{i}(1-\tau_{2})^{j}}
{\left(1-\gamma^2
   (1-\tau_{1})\right)^6 \left(1-\gamma^2 (1-\tau_{2})\right)^6} \label{final}
\end{equation}
where the matrix $A_{ij}(\g)$ is reported in \cite{supp}. The joint distribution of concurrence and squared norm is then given by
\begin{align}
P_{\gamma}(\bm{\mathcal{C}},\bm{\mathcal{N}})&=\Big\langle \delta\left(\bm{\mathcal{C}}-\frac{2\sqrt{\tau_1(1-\tau_1)\tau_2(1-\tau_2)}}{\tau_1+\tau_2-2\tau_1\tau_2}\right) \nonumber \\ &\phantom{=}\times\delta\left(\bm{\mathcal{N}}-\tau_1+\tau_2-2\tau_1\tau_2\right)\Big\rangle\label{jointCN}
\end{align}
where the average is taken with respect to the measure \eqref{final}. The final result is explicit but cumbersome and is confined to the attached Mathematica$\textsuperscript{\textregistered}$ notebook \cite{supp}. It is more instructive to analyze a few consequences of this calculation. First, the marginal distribution of concurrence alone, obtained by integrating $\bm{\mathcal{N}}$ out in \eqref{jointCN} has the surprisingly simple structure
\begin{equation}
P_\gamma(\bm{\mathcal{C}})=\sum_{k=0,4,6,8}c_k(\bm{\mathcal{C}})\gamma^k \label{finalC}
\end{equation}
where the coefficients $c_k(\bm{\mathcal{C}})$, expressed only in
terms of elementary functions, are listed explicitly in \cite{supp}. The limit $\g\to 0$, corresponding to ideal cavities,
yields 
\begin{equation}
P_\gamma(\bm{\mathcal{C}})=c_0(\bm{\mathcal{C}}), \label{idealC}
\end{equation}
recovering the probability of concurrence computed in
\cite{gopar}, Eq. (15). This is our first main result, which is shown in Fig. \ref{figprob} and analytically corroborates the numerical simulations presented in \cite{brasil}.

%\vspace{20pt}

Let us note that the simple polynomial
structure of \eqref{finalC}, not evident in \eqref{final}, is
essentially due to some symmetry of the integrand. In order to clarify
this point, it is instructive to consider the general $\gamma$-expansion of
\eqref{final}
\begin{equation}
P_{\gamma,\gamma}(\tau_{1},\tau_{2})=\sum_{k=0}^{\infty}\g^{2k} P^{(2k)}(\tau_{1},\tau_{2}). \label{expansion}
\end{equation}
Then, clearly $P^{(0)}(\tau_{1},\tau_{2})$ corresponds to the ideal case \cite{supp}, while $P^{(2)}(\tau_{1},\tau_{2})$ is antisymmetric under the
transformation $\tau_{1}\to 1- \tau_{1}$ and $\tau_{2}\to 1-
\tau_{2}$, while the concurrence is symmetric. This symmetry explains
why all moments of $\bm{\mathcal{C}}$ vanish up to the second order in
$\g$ and the first correction in \eqref{finalC} is
$\mathcal{O}(\g^{4})$. On the contrary, no trivial symmetry
kills all higher order terms and this enormous simplification can be
explained only by a non-linear change of variables~\cite{supp}.

\begin{figure}[htb]
\begin{center}
\includegraphics[scale=0.30,clip=true]{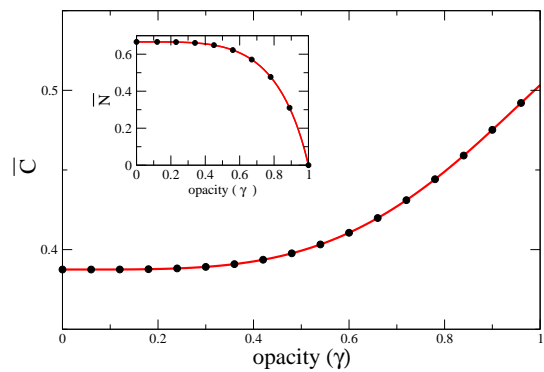}
\caption{ 
      Analytical prediction for the average concurrence $\overline{\bm{\mathcal{C}}}$ as a
      function of $\g$ (straight lines) together with numerical
      simulations (black circles). Inset: comparison between the average square norm $\overline{\bm{\mathcal{N}}}$ and numerical simulations.  \label{figaverage}}
\end{center}
\end{figure}

From \eqref{finalC}, the average concurrence reads
\begin{equation}
\overline{\bm{\mathcal{C}}} = a_{0}+a_{4} \g^4+a_{6} \g^6+a_{8} \g^8\label{avC}
\end{equation}
with coefficients $a_j$ listed in \cite{supp}. Being an increasing function of $\g$ (see Fig. \ref{figaverage}), Eq. \eqref{avC} leads to 
the somehow paradoxical conclusion that the average entanglement production is maximal for the completely ``opaque'' case $\g\to 1$ where no electrons travel across the cavity. This apparent paradox is resolved when one computes the average square norm of the entangled state from \eqref{jointCN}

\begin{equation}
\overline{\bm{\mathcal{N}}}=\small{\frac{2 \left(\gamma^2-1\right) \left(4 \left(\gamma^2-1\right)^2 \ln
   \left(1-\gamma^2\right)+\left(\gamma^4-6 \gamma^2+4\right)
   \gamma^2\right)}{\gamma^6}}\label{average_N}
\end{equation}
which is instead a (non-polynomial) \emph{decreasing} function of $\g$ (see inset in Fig. \ref{figaverage}). The simultaneous \emph{increase} in the average entanglement of the two-qubit state and \emph{decrease} in the average likelihood to detect it is the statistical analogue of the geometrical inequality $\bm{\mathcal{N}}(1+\bm{\mathcal{C}})<1$, first discovered for ideal leads \cite{novaes}. However, having now a tunable parameter (the opacity $\g$) at our disposal it is natural to ask whether the two competing effects above may be used to optimize the entanglement production process. We answer in the affirmative, considering the following natural observable
\begin{equation}
\overline{\left( \bm{\mathcal{C}}\vert \bm{\mathcal{N}}_0\right)}=\int_{0}^{1}d\bm{\mathcal{C}}\ \bm{\mathcal{C}} \int_{\bm{\mathcal{N}}_0}^{1}d \bm{\mathcal{N}}\ P_{\gamma}(\bm{\mathcal{C}},\bm{\mathcal{N}})\label{CAC}
\end{equation}
namely the \emph{constrained} average concurrence under the requirement of a minimal detectable norm $\bm{\mathcal{N}}_0$.

\begin{figure}[htb]
\centerline{\epsfig{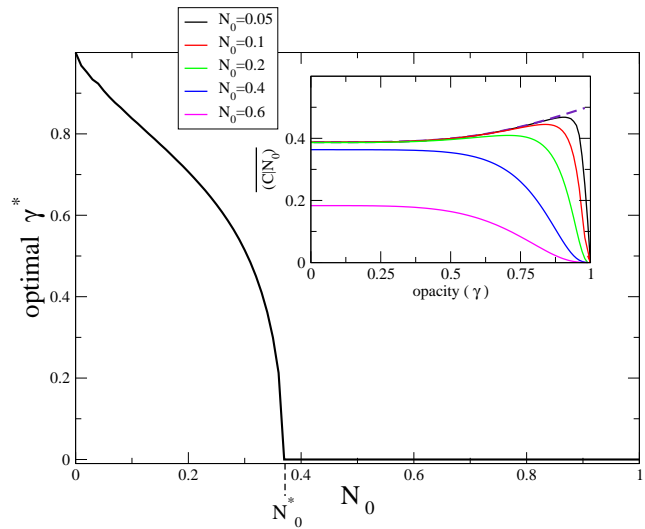}} \caption{Behavior of the optimal $\gamma^\star$ as a function of the minimal required squared norm $\bm{\mathcal{N}}_0$. When $\bm{\mathcal{N}}_0$ is greater than $\bm{\mathcal{N}}_0^\star\simeq 0.3693\dots$ the maximal average entanglement production is reached for the case of ideal leads. Inset: distribution of $\overline{\left( \bm{\mathcal{C}}\vert \bm{\mathcal{N}}_0\right)}$ for different values of $\bm{\mathcal{N}}_0$, where the bifurcation effect around $\gamma=0$ is now explicit. \label{fig:optimal}}
\end{figure}

The quantity \eqref{CAC} has a rich and intriguing behavior as a function of $\g$. In particular, it develops a maximum for a nonzero value $\g^\star( \bm{\mathcal{N}}_0)$ as long as $ \bm{\mathcal{N}}_0$ is below a critical value $\bm{\mathcal{N}}_0^\star\simeq 0.3693\dots$, indicating that for a given value of the minimal detectable norm the average entanglement production can be maximized by finely tuning the opacity of the non-ideal lead. However, if a ``too high'' experimental resolution is required, the optimal option is to stick to the ideal case. This behavior is summarized in Fig. \ref{fig:optimal} and arises as a consequence of an interesting bifurcation effect: the value $\bm{\mathcal{N}}_0^\star$ indeed marks the transition from a local minimum to a global maximum of the function $\overline{\left( \bm{\mathcal{C}}\vert \bm{\mathcal{N}}_0\right)}$
around $\gamma=0$ as it can be derived from a local stability
analysis~\cite{supp}. This prediction constitutes the second main result of
our work.

In order to check the analytical predictions, we use the same simple
numerical algorithm described in detail in Ref. \cite{brasil}. The
$4\times 4$ scattering matrix of the non-ideal system
$\bm{\mathcal{S}}$ can be decomposed as $
\bm{\mathcal{S}}=\bm{\mathcal{R}} +\bm{\mathcal{T}}
(\mathbf{1}-\bm{\mathcal{S}}_0\bm{\mathcal{R}})^{-1}
\bm{\mathcal{S}}_0 \bm{\mathcal{T}} $ where $\bm{\mathcal{S}}_0$ is a
random matrix distributed uniformly within the unitary group and
represents the scattering matrix of an ideal cavity.  The matrices
$\bm{\mathcal{R}} $ and $\bm{\mathcal{T}} $ include information on the
transparency of the contacts, as follows:
\begin{align}
\bm{\mathcal{R}} &= \mathrm{diag}(\mathrm{i}\g,\mathrm{i}\g,0,0)\\
\bm{\mathcal{T}}  &= \mathrm{diag}(\sqrt{1-\g^2},\sqrt{1-\g^2},1,1)
\end{align}

We generate $\sim 10^6$ matrices $\bm{\mathcal{S}}_0 $ using the
QR-based algorithm described in \cite{edelmanrao}, and we build the
corresponding matrices $\bm{\mathcal{S}}$. Having extracted the
submatrix $\mathbf{t}$ and diagonalized $\mathbf{t t}^\dagger$, we
collect its eigenvalues $\tau_{1}$ and $\tau_2$ and from them we
construct the concurrence $\bm{\mathcal{C}}$ and the squared norm
$\bm{\mathcal{N}}$. Our numerical simulations are all in perfect agreement with our theoretical results.

In summary, we have investigated analytically the distributions of
concurrence and squared norm for the entanglement production in a chaotic quantum dot
supporting one ideal and one non-ideal lead. The concurrence distribution 
is in perfect agreement with the numerical
findings in \cite{brasil} (Fig. 6 bottom) and recovers, in the proper
limit, the known results on the ideal transparency
case~\cite{gopar}. Geometrical constraints on the entanglement production are responsible for
competing effects in the behavior of the average values $\overline{\bm{\mathcal{C}}}$ and $\overline{\bm{\mathcal{N}}}$ as a function of the opacity $\g$, which can be exploited to maximize the \emph{constrained} average concurrence when a minimal detectable norm $\bm{\mathcal{N}}_0$ is required. In particular, we confirm that the entanglement production is enhanced when the transparency of one of the contacts is not ideal, but only up to a critical value $\bm{\mathcal{N}}_0^\star\simeq 0.3693\dots$ beyond which no net improvement with respect to the ideal case can be exploited.

  This study lays the groundwork for future research directions. Given that the structure of the probability distribution of concurrence as a function of the opacity is simple and elegant, this clearly hints
  towards a deeper connection with physical symmetries of the
  system. Moreover, we showed that a bifurcation phenomenon arises when
the optimal entanglement production is studied in a non-ideal setting. These theoretical
predictions may be tested experimentally and the comprehension of their
  generality when, for instance, both leads are
  non-ideal, or in other multi-channel experiments, is still
  lacking and represents a challenging issue whose solution is much
  called for.

%\vspace{20pt}

%  and we have obtained also the analytical expression of
% the first moments, which may be tested experimentally. 

\textit{Acknowledgments:} We acknowledge the warm hospitality
at the conference 'Non-equilibrium fluctuation-response relations' at
Giglio Island, where this work was initiated. We also thank Carlo
Barbieri of Wolfram Alpha for useful discussions on some Mathematica routines and Markus B\"uttiker for helpful correspondence.

%%%%%%%%%%%%%%%%%%%%%%%%%%%%%%%%%%%%%%%
%      BIBLIOGRAPHY
%%%%%%%%%%%%%%%%%%%%%%%%%%%%%%%%%%%%%%%

\begin{widetext}
\section*{\LARGE Supplemental Material}
In this section we have collected the material that could not be
included in the main text, providing technical details and derivations. We also attach a Mathematica$\textsuperscript{\textregistered}$ notebook with all relevant formulae and the code to produce numerical simulations.\\

In section~\ref{app:1} we discuss the probability measure to impose on the
scattering matrix as well as the joint probability density of transmission eigenvalues
in the ideal and non-ideal cases. In
section~\ref{app:2} one of the main result of the Letter is derived,
namely the exact distribution of the concurrence and its polynomial
structure. Finally, in section~\ref{app:3} the main issue of
the joint probability distribution of concurrence and squared norm is
addressed, together with the expansion for small $\gamma$ which
confirms the bifurcation phenomena leading to the appearance of the optimal $\gamma^\star$.\\

\section{Probability measure on the scattering matrix \label{app:1}}
The transport properties of the cavity depicted in Fig.~1 of the main
text are governed by the probability measure to impose on the scattering matrix $\bm{\mathcal{S}}$. It is well
established that:
\begin{itemize}
\item in the case of \emph{ideal} leads, the distribution of
  $\bm{\mathcal{S}}$ is \emph{uniform} within the unitary group: in
  the presence of TRI, the statistics of an ensemble of unitary and
  symmetric $\bm{\mathcal{S}}$ matrices is described by the so-called
  Circular Orthogonal Ensemble ($\beta=1$). When TRI is absent, the
  statistical properties of $\bm{\mathcal{S}}$ are described by the
  Circular Unitary Ensemble ($\beta=2$).  The uniformity of
  $\bm{\mathcal{S}}$ induces a non-trivial joint density of the
  transmission eigenvalues $\tau_{1,2}$ of the transmission matrix
  $\mathbf{t t}^\dagger$ \cite{jp1,baranger-mello,jalabert}, given by
  the Jacobi ensemble of random matrices:
\begin{equation}
\label{dist_taus}
P_{0,0}( \tau_1,\tau_2)=c_\beta| \tau_1 - \tau_2 |^{\beta}
(\tau_1 \tau_2)^{\beta/2-1},
\end{equation}
where $c_1=3/4$ and $c_2=6$ are normalization constants.

\item in the case of \emph{non-ideal} leads supporting $n_{\rm L}$
  (left) and $n_{\rm R}$ (right) propagating channels, the
  distribution of $\bm{\mathcal{S}}$ is instead given by the so-called
  Poisson kernel \cite{brouw},
\begin{equation}
\label{pk}
    P_\beta({\bm {\mathcal S}}) \propto \big[
    {\rm det}(
        \openone_N -  \bar{\bm {\mathcal S}} {\bm {\mathcal S}}^\dagger)\,
        {\rm det}(
        \openone_N -   {\bm {\mathcal S}} \bar{\bm {\mathcal S}}^\dagger)
    \big]^{ \beta/2 -1  -  \beta N/2}.
\end{equation}
\end{itemize}
The $N=n_{\rm L}+n_{\rm R}$ eigenvalues $\hat{\bm \gamma} = {\rm diag}
(\{\sqrt{1-\Gamma_j}\})$ of the average matrix $\bar{\bm {\mathcal
    S}}$ characterize couplings between the cavity and the leads in
terms of tunnel probabilities $\Gamma_j$ of the $j$-th electron
channel in the leads. In the ideal limit $\Gamma_i\to 1$, we recover
uniformity of $\bm{\mathcal{S}}$ within the unitary group. In contrast
with the ideal case, however, no general derivation exists of the
joint density of transmission eigenvalues in the non-ideal
case. However, for $\beta=2$ in the special case where the left lead
supports $n_{\rm L}$ propagating modes characterized by a set of
tunnel probabilities $\hat{{\bm \Gamma}}_{\rm
  L}=(\Gamma_1,\dots,\Gamma_{n_{\rm L}})$, while the right lead
(supporting $n_{\rm R} \ge n_{\rm L}$ open channels) is kept ideal so
that $\hat{{\bm \Gamma}}_{\rm R}=(\Gamma_{n_{\rm
    L}+1},\dots,\Gamma_{n_{\rm R}})=\openone_{n_{\rm R}-n_{\rm L}}$,
the joint probability density function $P_{(\hat{{\bm \gamma}}_{\rm
    L}|\,{\bm 0})}({\bm R})$ of reflection eigenvalues $\{R_j=1-\tau_j
\}$ equals \cite{vidal}

\begin{equation}\label{jpdf-1}
   P_{(\hat{{\bm \gamma}}_{\rm L}|\,{\bm 0})}(R_1,\dots,R_{n_{\rm L}})  \propto
    \, \prod_{j<k}(R_j-R_k) \,
            {\rm det}\Big[
                {}_2 F_1 (n_{\rm R}+1,n_{\rm R}+1; 1;\, \gamma_j^2 R_k)
            \Big]_{(j,k)\in (1, n_{\rm L})} \prod_{j=1}^{n_{\rm L}} (1-R_j)^{\nu}.
\end{equation}

where $\nu=|n_{\rm L}-n_{\rm R}|$ and $\{\g_1^2
=1-\Gamma_1,\ldots,\g_{n_{\rm L}}^2=1-\Gamma_{n_{\rm L}}\}$ is a set
of $n_{\rm L}$ opacity parameters related to the associated tunnel
probabilities in the non-ideal leads, whilst ${}_p F_q$ is the Gauss
hypergeometric function.

It is now necessary to specialize Eq.~\eqref{jpdf-1} to the case of two
transmission eigenvalues $(n_{\rm L}=n_{\rm R}=2)$ and for
$\g_1=\g_2=\g$, since both channels in the left lead clearly need to
have the same opacity. This leads after lengthy simplifications to Eq. (4) of the Letter.

We give here the form of the matrix $A_{ij}(\g)$ in the expansion of the joint probability density of eigenvalues $P_{\g,\g}(\tau_1,\tau_2)$ (Eq. (4) of the Letter).

\begin{equation}
A_{ij}(\g)=\left(
\begin{array}{ccccc}
 0 & 0 & 6 & 12 \g^2 & 2 \g^4 \\
 0 & -12 & -12 \g^2 & 52 \g^4 & 12 \g^6 \\
 6 & -12 \g^2 & -108 \g^4 & -12 \g^6 & 6 \g^8 \\
 12 \g^2 & 52 \g^4 & -12 \g^6 & -12 \g^8 & 0 \\
 2 \g^4 & 12 \g^6 & 6 \g^8 & 0 & 0 \\
\end{array}
\right)
\end{equation}

\section{Distribution of the concurrence and polynomial structure\label{app:2}}
We now start from Eq. (4) of the Letter for the joint probability density of transmission eigenvalues:

\begin{equation}
P_{\gamma,\gamma}(\tau_{1},\tau_{2})=\frac{(\gamma^{2}-1)^{8} \sum_{i,j=0}^{4}A_{ij}(\g)(1-\tau_{1})^{i}(1-\tau_{2})^{j}}
{\left(1-\gamma^2
   (1-\tau_{1})\right)^6 \left(1-\gamma^2 (1-\tau_{2})\right)^6} \label{finalapp}
\end{equation}

The marginal distribution of concurrence is given by the following integral:
\begin{equation}
P_\gamma(\bm{\mathcal{C}})=\int_{[0,1]^2}d\tau_1 d\tau_2 P_{\gamma,\gamma}(\tau_{1},\tau_{2})\delta\left(\bm{\mathcal{C}}-\frac{2\sqrt{\tau_1 (1-\tau_1)\tau_2 (1-\tau_2)}}{\tau_1+\tau_2-2\tau_1\tau_2}\right)
\end{equation}

We make the change of variables (valid for $\g\neq 1$):
\begin{equation}
\begin{cases}
(1-\g ^2 )z_1 &=\frac{\tau_{1}}{1-\tau_1}\\
(1-\g ^2 )z_2 &=\frac{\tau_{2}}{1-\tau_2}\\
\end{cases} \label{change}
\end{equation}

This way, after simplifications, the integral becomes:
\begin{equation}
P_\gamma(\bm{\mathcal{C}})=\int_{[0,\infty]^2}dz_1 dz_2 \left[\sum_{k=0,2,4,6,8}\g^k f_k(z_1,z_2)\right]\delta\left(\bm{\mathcal{C}}-2\frac{\sqrt{z_1 z_2}}{z_1+z_2}\right).\label{Pc}
\end{equation}

Note that the change of variable~\eqref{change} is essential in bringing the polynomial structure in $\g$ to the surface. The functions involved in Eq.~\eqref{Pc} are:

\begin{align}
f_0(z_1,z_2) &=\frac{6 (z_1 - z_2)^2}{(1 + z_1)^4 (1 + z_2)^4}\\
f_2(z_1,z_2) &= -\frac{24 (z_1 - z_2)^2 (-1 + z_1 z_2)}{(1 + z_1)^5 (1 + z_2)^5}\\
f_4(z_1,z_2) &= \frac{4 (z_1 - z_2)^2 (15 + z_1 (6 - 4 z_2) + 6 z_2 - 4 z_2^2 + 
   z_1^2 (-4 + 9 z_2^2)}{(1 + z_1)^6 (1 + z_2)^6}\\
   f_6(z_1,z_2) &= -\frac{8 (z_1 - z_2)^2 (-3 - (-6 + z_2) z_2 + z_1 (6 + (8 - 3 z_2) z_2) + 
    z_1^2 (-1 + 3 (-1 + z_2) z_2)}{(1 + z_1)^6 (1 + z_2)^6}\\
    f_8(z_1,z_2) &=\frac{2 (z_1 - z_2)^2 (3 + (-6 + z_2) z_2 + z_1^2 (1 + 3 (-2 + z_2) z_2) + 
   z_1 (-6 + 28 z_2 - 6 z_2^2)}{(1 + z_1)^6 (1 + z_2)^6}
\end{align}

The integrals of the form:
\begin{equation}
c_k(\bm{\mathcal{C}})=\int_{[0,\infty]^2}dz_1 dz_2  f_k(z_1,z_2) \delta\left(\bm{\mathcal{C}}-2\frac{\sqrt{z_1 z_2}}{z_1+z_2}\right)
\end{equation}

can be computed by first expanding the delta function as:
\begin{align}
\nonumber\delta\left(\bm{\mathcal{C}}-2\frac{\sqrt{z_1 z_2}}{z_1+z_2}\right) &=\frac{(\phi^{(+)}(\bm{\mathcal{C}})+1)^2 \sqrt{\phi^{(+)}(\bm{\mathcal{C}})}}{\phi^{(+)}(\bm{\mathcal{C}})-1}z_2\ \delta(z_1-\phi^{(+)}(\bm{\mathcal{C}})z_2)+\\
&\frac{(\phi^{(-)}(\bm{\mathcal{C}})+1)^2 \sqrt{\phi^{(-)}(\bm{\mathcal{C}})}}{1-\phi^{(-)}(\bm{\mathcal{C}})}z_2\ \delta(z_1-\phi^{(-)}(\bm{\mathcal{C}})z_2)
\end{align}
where
\begin{equation}
\phi^{(\pm)}(\bm{\mathcal{C}})=\frac{2-\bm{\mathcal{C}}^2\pm 2\sqrt{1-\bm{\mathcal{C}}^2}}{\bm{\mathcal{C}}^2}
\end{equation}

yielding eventually:
\begin{equation}
c_k(\bm{\mathcal{C}})=\frac{(\phi^{(+)}(\bm{\mathcal{C}})+1)^2 \sqrt{\phi^{(+)}(\bm{\mathcal{C}})}}{\phi^{(+)}(\bm{\mathcal{C}})-1} c_k^{(+)}(\bm{\mathcal{C}})
+\frac{(\phi^{(-)}(\bm{\mathcal{C}})+1)^2 \sqrt{\phi^{(-)}(\bm{\mathcal{C}})}}{1-\phi^{(-)}(\bm{\mathcal{C}})}c_k^{(-)}(\bm{\mathcal{C}})
\end{equation}

where:
\begin{equation}
c_k^{(\pm)}(\bm{\mathcal{C}}) =\int_0^\infty dz\ z\ f_k\left(\phi^{(\pm)}(\bm{\mathcal{C}}) z,z\right)\label{ckplus}
\end{equation}

The single integrals in \eqref{ckplus} can be computed and after lengthy algebra we eventually get to the following coefficients:

\begin{align}
c_0(\bm{\mathcal{C}}) &=\frac{\bm{\mathcal{C}} \left(4 \sqrt{
    1 - \bm{\mathcal{C}}^2} (11 + 4\ \bm{\mathcal{C}}^2) + (6 + 
      9\ \bm{\mathcal{C}}^2) \ln\left(\frac{2-\bm{\mathcal{C}}^2-2
      \sqrt{1-\bm{\mathcal{C}}^2}}{2-\bm{\mathcal{C}}^2+2\sqrt{1-\bm{\mathcal{C}}^2}}\right)\right)}{2 (-1 + \bm{\mathcal{C}}^2)^3}\\
 \nonumber c_4(\bm{\mathcal{C}}) &= \frac{\bm{\mathcal{C}}}{3(\bm{\mathcal{C}}^2-1)^4}\left( 4 \sqrt{1-\bm{\mathcal{C}}^2} (62+221\ \bm{\mathcal{C}}^2+32\ \bm{\mathcal{C}}^4)+6 (8+64\ \bm{\mathcal{C}}^2+33\ \bm{\mathcal{C}}^4)\times\right.\\
 &\left. \times\mathrm{arctanh}\left(\frac{2\sqrt{1-\bm{\mathcal{C}}^2}}{\bm{\mathcal{C}}^2-2}\right)\right)\\
 \nonumber c_6(\bm{\mathcal{C}}) &=-\frac{\bm{\mathcal{C}}}{6(\bm{\mathcal{C}}^2-1)^4}\left(4\sqrt{1-\bm{\mathcal{C}}^2} (62+221 \bm{\mathcal{C}}^2+32\ \bm{\mathcal{C}}^4)+\right. \\
  &\left. 3 (8+64\ \bm{\mathcal{C}}^2+33\ \bm{\mathcal{C}}^4) \ln \left(\frac{2-\bm{\mathcal{C}}^2-2\sqrt{1-\bm{\mathcal{C}}^2}}{2-\bm{\mathcal{C}}^2+2\sqrt{1-\bm{\mathcal{C}}^2}}\right)\right)\\
\nonumber  c_8(\bm{\mathcal{C}}) &=\frac{\bm{\mathcal{C}}}{12(\bm{\mathcal{C}}^2-1)^5}\left(2\sqrt{1-\bm{\mathcal{C}}^2}(80+1212\ \bm{\mathcal{C}}^2+1431\ \bm{\mathcal{C}}^4+112\ \bm{\mathcal{C}}^6)+\right. \\
  &\left. 3(8+236\ \bm{\mathcal{C}}^2+554\ \bm{\mathcal{C}}^4+147\ \bm{\mathcal{C}}^6)\mathrm{arctanh}\left(\frac{2\sqrt{1-\bm{\mathcal{C}}^2}}{\bm{\mathcal{C}}^2-2}\right)\right)
\end{align}
Note that the coefficient $c_2(\bm{\mathcal{C}})$ is identically zero. Given the aforementioned symmetries, it is straightforward to observe that:
\begin{equation}
\left<\bm{\mathcal{C}}^{n}\right>=\int d\tau_1d\tau_2 P^{(0)}(\tau_{1},\tau_{2}) \left[\frac{2\sqrt{\tau_1(1-\tau_1)\tau_2(1-\tau_2)}}{\tau_1+\tau_2-2\tau_1\tau_2}\right]^{n}\equiv 0
\end{equation}
for all values of $n$. Note that also for higher order terms than $\mathcal{O}(\g^{8})$, all the integrals of the kind:
\begin{equation}
\left<\bm{\mathcal{C}}^{n}\right>=\int d\tau_1d\tau_2 P^{(2k)}(\tau_{1},\tau_{2}) \left[\frac{2\sqrt{\tau_1(1-\tau_1)\tau_2(1-\tau_2)}}{\tau_1+\tau_2-2\tau_1\tau_2}\right]^{n}
\end{equation}
vanish for all $k>4$ and every $n$ ,  as can be clearly deduced
from the exact probability density of the concurrence $\bm{\mathcal{C}}$. However, quite interestingly, the simple symmetry
argument given above or other simple transformations do not hold for these terms.

In conclusion, moments of arbitrary order can be computed exactly if needed. For instance, the first two are:
\begin{align}
\overline{\bm{\mathcal{C}}}&= a_{0}+a_{4} \g^4+a_{6} \g^6+a_{8} \g^8 \\
\overline{\bm{\mathcal{C}}^{2}}&= b_{0}+b_{4} \g^4+b_{6} \g^6+b_{8} \g^8 
\end{align}

\begin{align}
a_{0} &= \frac{1}{16} \pi  (21 \pi -64) \\
a_{4} &=\frac{1}{48} \pi  (327 \pi -1024)\\  
a_{6} &=-\frac{1}{96} \pi  (327 \pi -1024) \\ 
a_{8} &=\frac{\pi  (9129 \pi -28672)}{3072} \\
b_{0} &=-22+\frac{9\pi^{2}}{4}\\
b_{4} &=\frac{1}{6}\left(-976+99\pi^{2} \right) \\
b_{6} &=\frac{244}{3}-\frac{33 \pi ^2}{4}\\
b_{8} &=\frac{147 \pi ^2}{16}-\frac{272}{3}
\end{align}

\section{Joint distribution of concurrence and squared norm\label{app:3}}

The joint distribution $P_\g(\bm{\mathcal{N}},\bm{\mathcal{C}})$ is given by
\begin{equation}
P_{\gamma}(\bm{\mathcal{C}},\bm{\mathcal{N}})=\Big\langle \delta\left(\bm{\mathcal{C}}-\bm{\mathcal{C}}_{def}\right)\delta\left(\bm{\mathcal{N}}-\bm{\mathcal{N}}_{def}\right)\Big\rangle\label{jointCNsupp}
\end{equation}
where
\begin{equation}
\begin{cases}
\bm{\mathcal{C}}_{def}&=\frac{2\sqrt{\tau_1(1-\tau_1)\tau_2(1-\tau_2)}}{\tau_1+\tau_2-2\tau_1\tau_2}\\
\bm{\mathcal{N}}_{def}&=\tau_1+\tau_2-2\tau_1\tau_2\label{sys_of_eq}
\end{cases}
\end{equation}
The distribution can be computed generalizing the
procedure used in~\cite{novaes} for the ideal case. The general
solution can then be written in this form

 \begin{equation}
   P_{\gamma}(\bm{\mathcal{C}},\bm{\mathcal{N}})= 4\Theta(1-\bm{\mathcal{N}}(1+\bm{\mathcal{C}}))\vert\mathcal{J}(\bm{\mathcal{C}},\bm{\mathcal{N}})\vert \sum_{i=1}^{4}  P_{\gamma}(\tau^{*}_{1,i} (\bm{\mathcal{C}},\bm{\mathcal{N}}),\tau^{*}_{2,i} (\bm{\mathcal{C}},\bm{\mathcal{N}})).
 \end{equation}
 where $\Theta(\dots)$ is the Heaviside step function, the
 $\tau^{*}_{k,i}$ are the solutions of the system of equations~\eqref{sys_of_eq}
 and $\mathcal{J}(\bm{\mathcal{C}}, \bm{\mathcal{N}})$ is the Jacobian of the
 transformation. The Heaviside function denotes the part of the plane
 $(\bm{\mathcal{N}},\bm{\mathcal{C}})$ where the system admits four solutions.
 
Observing the system~\eqref{sys_of_eq}, one concludes that the four solutions enjoy the channel-to-channel symmetry:
\begin{equation}
\begin{cases}
\tau_1 &\to \tau_2\\
\tau_2 &\to \tau_1\label{ccsymm}
\end{cases}
\end{equation}

and the transmission-to-reflection symmetry:

\begin{equation}
\begin{cases}
\tau_1 &\to 1-\tau_1\\
\tau_2 &\to 1-\tau_2\label{rlsymm}
\end{cases}
\end{equation}

In the general case the symmetry \eqref{rlsymm} is broken, whereas the symmetry \eqref{ccsymm} can be exploited, yielding
 \begin{equation}
   P_{\gamma}(\bm{\mathcal{C}},\bm{\mathcal{N}})= 2\Theta(1-\bm{\mathcal{N}}(1+\bm{\mathcal{C}}))\vert\mathcal{J}(\bm{\mathcal{C}},\bm{\mathcal{N}}) \vert \left[P_{\gamma}(\tau^{*}_{1} (\bm{\mathcal{C}}, \bm{\mathcal{N}}),\tau^{*}_{2} (\bm{\mathcal{C}}, \bm{\mathcal{N}}))+ P_{\gamma}(1-\tau^{*}_{1} (\bm{\mathcal{C}}, \bm{\mathcal{N}}),1-\tau^{*}_{2} (\bm{\mathcal{C}}, \bm{\mathcal{N}}))\right] .\label{joint}
\end{equation}
where $(\tau^{*}_{1},\tau^{*}_{2})$ is simply any of the four solutions of \eqref{sys_of_eq}.

 On the contrary, in the limit of the ideal case the
 symmetry~\eqref{rlsymm} is restored and the expression is further simplified:

 \begin{equation}
   P_{0}(\bm{\mathcal{C}},\bm{\mathcal{N}})= 4\Theta(1-\bm{\mathcal{N}}(1+\bm{\mathcal{C}}))\vert\mathcal{J}(\bm{\mathcal{C}},\bm{\mathcal{N}}) \vert P_{0}(\tau^{*}_{1} (\bm{\mathcal{C}}, \bm{\mathcal{N}}),\tau^{*}_{2} (\bm{\mathcal{C}}, \bm{\mathcal{N}})).
 \end{equation}

 A similar situation occurs in the study of the distribution of the
 squared norm, making the resulting expression for the non-ideal case
 highly involved (see the attached
 Mathematica$\textsuperscript{\textregistered}$ notebook) when compared to the case $\gamma=0$.

\subsection{Local stability analysis}
The expression~\eqref{joint} can be used to probe the behavior of the optimal
value $\gamma^{\star}$ described in the Letter, and in particular the
bifurcation effect. It is sufficient
to expand $P_{\gamma}(\bm{\mathcal{C}},\bm{\mathcal{N}})$ around the ideal $\gamma\to 0$ case

\begin{equation}
P_{\gamma}(\bm{\mathcal{C}},\bm{\mathcal{N}})=P^{(0)}(\bm{\mathcal{C}},\bm{\mathcal{N}})+P^{(4)}(\bm{\mathcal{C}},\bm{\mathcal{N}})\gamma^{4}+ \mathcal{O}(\gamma^{6})
\end{equation}

The term $P^{(2)}(\bm{\mathcal{C}},\bm{\mathcal{N}})$ vanishes for the symmetry already encountered for the marginal distribution of concurrence. The two coefficients in the expansion can be derived analytically and read
\begin{eqnarray}
P^{(0)}(\bm{\mathcal{C}},\bm{\mathcal{N}}) &=&  \Theta
   (1-(\bm{\mathcal{C}}+1)\bm{\mathcal{N}}) \frac{12 \bm{\mathcal{N}}^3 \bm{\mathcal{C}}\sqrt{1-\bm{\mathcal{C}}^2}}{\sqrt{1-2\bm{\mathcal{N}}+\bm{\mathcal{N}}^2 (1-\bm{\mathcal{C}}^2)}}\label{noveq}\\
P^{(4)}(\bm{\mathcal{C}},\bm{\mathcal{N}}) &=& \frac{16 \bm{\mathcal{C}} \bm{\mathcal{N}}^3
   \left(25 \bm{\mathcal{C}}^4
   \bm{\mathcal{N}}^2-50 \bm{\mathcal{C}}^2
   \bm{\mathcal{N}}^2+51 \bm{\mathcal{C}}^2
   \bm{\mathcal{N}}-24 \bm{\mathcal{C}}^2+25
   \bm{\mathcal{N}}^2-51
   \bm{\mathcal{N}}+24\right) \Theta
   (1-(\bm{\mathcal{C}}+1)
   \bm{\mathcal{N}})}{\sqrt{\left(\bm{\mathcal{C}}^2-1\right)
   \left(\bm{\mathcal{C}}^2
   \bm{\mathcal{N}}^2-\bm{\mathcal{N}}^2+2
   \bm{\mathcal{N}}-1\right)}}\label{expansion}
\end{eqnarray}

Note that \eqref{noveq} is obviously in agreement with the ideal-case result in \cite{novaes}, Eq. 14.

From \eqref{expansion} one defines the following function, whose sign determines the concavity of the function $\overline{(\bm{\mathcal{C}}\vert \bm{\mathcal{N}}_{0})}$ at $\gamma=0$
\begin{equation}
\chi(\bm{\mathcal{N}}_{0})=\int_{0}^{1}d\bm{\mathcal{C}}\ \bm{\mathcal{C}}\int_{\bm{\mathcal{N}}_{0}}^{1} d \bm{\mathcal{N}}\ P^{(4)}(\bm{\mathcal{C}},\bm{\mathcal{N}})
\end{equation}

It is positive for $\bm{\mathcal{N}}_{0}<\bm{\mathcal{N}}_0^{\star}$ and negative for $\bm{\mathcal{N}}_{0}>\bm{\mathcal{N}}_0^{\star}$ (see Fig. \ref{curvature}), where $\overline{(\bm{\mathcal{C}}\vert \bm{\mathcal{N}}_{0})}$ at $\gamma=0$ has a global maximum. 
\end{widetext}

\begin{figure}[htb]
  \centerline{\epsfig{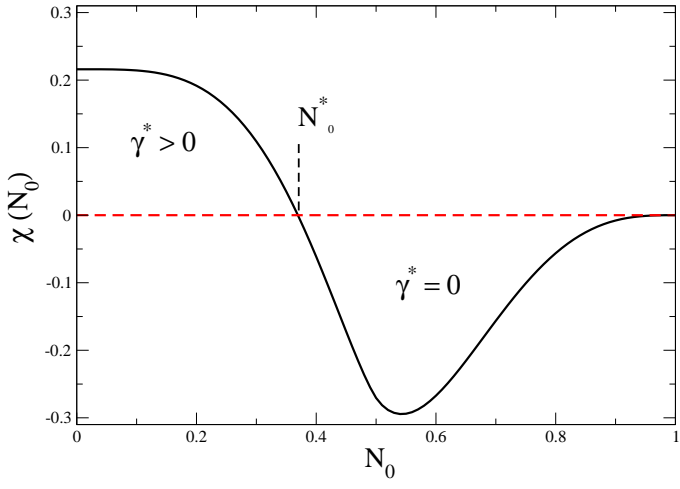}} 
  \caption{Plot of the function $\chi(\bm{\mathcal{N}}_{0})$.}\label{curvature}
\end{figure}


\begin{thebibliography}{30}

\bibitem{EPR35} A. Einstein, B. Podolsky, and N. Rosen,
			 Phys. Rev. {\bf 47}, 777 (1935).

\bibitem{NC-book} M. A. Nielsen and I. L. Chuang, {\it Quantum Computation and Quantum Information}  
(Cambridge University Press, Cambridge, 2000).

\bibitem{alber} G. Alber, T. Beth, P. Horodecki, R. Horodecki, M. R\"otteler, H. Weinfurter, R. Werner, and A. Zeilinger, \textit{Quantum Information} (Springer, Berlin, 2001), Vol. 173.

\bibitem{B06} C. W. J. Beenakker, 
		     in {\it Quantum Computers, Algorithms and Chaos}, 
		     Proceedings of the International School of Physics "Enrico Fermi," Varenna, 2005, 
		     (IOS Press, Amsterdam, 2006).
		     
\bibitem{B07} G. Burkard, J. Phys.: Condens. Matter {\bf 19}, 233202 (2007).

\bibitem{inter}
P. Recher, E. V. Sukhorukov, and D. Loss, Phys. Rev. B {\bf 63}, 165314 (2001); 
G. B. Lesovik, T. Martin, and G. Blatter, Eur. Phys. J. B {\bf 24}, 287 (2001);
W. D. Oliver, F. Yamaguchi, and Y. Yamamoto, Phys. Rev. Lett. {\bf 88}, 037901 (2002);  
V. Bouchiat, N. Chtchelkatchev, D. Feinberg, G. B. Lesovik, T. Martin, and J. Torres, Nanotechnology {\bf 14}, 77 (2003); 
D. S. Saraga and D. Loss, Phys. Rev. Lett. {\bf 90}, 166803 (2003). 

\bibitem{noninter}
S. Bose and D. Home, Phys. Rev. Lett. {\bf 88}, 050401 (2002);
C. W. J. Beenakker, C. Emary, M. Kindermann, and J. L. van Velsen, Phys. Rev. Lett. {\bf 91}, 147901 (2003); 
P. Samuelsson, E. V. Sukhorukov, and M. B\"uttiker, Phys. Rev. Lett. {\bf 92}, 026805 (2004) and New. J. Phys. {\bf 7}, 176 (2005); 
A. I. Signal and U. Z\"ulicke, Appl. Phys. Lett. {\bf 87}, 102102 (2005); 
A. V. Lebedev, G. B. Lesovik, and G. Blatter, Phys. Rev. B {\bf 71}, 045306 (2005).

\bibitem{FMF06}  D. Frustaglia, S. Montangero, and R. Fazio, 
			  Phys. Rev. B {\bf 74}, 165326 (2006).
 
\bibitem{BKMY04} C. W. J. Beenakker, M. Kindermann, C. M. Marcus, and A. Yacoby, 
			    in {\it Fundamental Problems of Mesoscopic Physics}, edited by I. V. Lerner, B. L. Altshuler, 
			    and Y. Gefen, NATO Science Series II vol. 154 (Kluwer, Dordrecht, 2004).


\bibitem{samuelsson} An orbital entangler for electrons based on an interacting mechanism was first introduced in P. Samuelsson, E. V. Sukhorukov, and M. B\"uttiker, Phys. Rev. Lett. {\bf 91}, 157002 (2003).

\bibitem{cond} F. Mezzadri and N. J. Simm, J. Math. Phys. {\bf 53}, 053504 (2012) and \textit{ibid.} {\bf 52}, 103511 (2011); V. Al. Osipov and E. Kanzieper, Phys. Rev. Lett. {\bf 101}, 176804 (2008);
B. A. Khoruzhenko, D. V. Savin, and H.-J. Sommers, Phys. Rev. B {\bf 80}, 125301 (2009); H.-J. Sommers, W. Wieczorek, and D. V. Savin, Acta Phys. Polon. A {\bf 112}, 691 (2007); M. Novaes,
Phys. Rev. B {\bf 78}, 035337 (2008) and \textit{ibid.} {\bf 75}, 073304 (2007); P. Vivo, S. N. Majumdar, and O. Bohigas, Phys. Rev. Lett. {\bf 101}, 216809 (2008) and Phys. Rev. B {\bf 81}, 104202 (2010);
P. W. Brouwer and C. W. J. Beenakker, Phys. Rev. B {\bf 50}, 11263 (1994); C. W. J. Beenakker,
Rev. Mod. Phys. \textbf{69}, 731 (1997);  P. A. Mello and N. Kumar, {\it Quantum Transport in
  Mesoscopic Systems. Complexity and statistical fluctuations} (Oxford
  University Press, Oxford, 2004).

\bibitem{gopar} V. A. Gopar and D. Frustaglia, Phys. Rev. B {\bf 77}, 153403 (2008). 

\bibitem{novaes} S. Rodriguez-P\'erez and M. Novaes, Phys. Rev. B {\bf 85}, 205414 (2012).

\bibitem{butt} P. Samuelsson, I. Neder, and M. B\"uttiker, Physica Scripta {\bf 2009}, 014023 (2009). 

\bibitem{brasil} F. A. G. Almeida and A. M. C. Souza, Phys. Rev. B {\bf 82}, 115422 (2010).

\bibitem{almeidamacedo} F. A. G. Almeida, S. Rodriguez-P\'erez, and A. M. S. Mac\^edo, Phys. Rev. B {\bf 80}, 125320 (2009).



\bibitem{vidal}  P. Vidal and E. Kanzieper, Phys. Rev. Lett. {\bf 108}, 206806 (2012). 

\bibitem{supp} See Supplemental Material.

\bibitem{W98} W. K. Wootters, 
	               Phys. Rev. Lett. {\bf 80}, 2245 (1998).
	               
\bibitem{jp1} R. A. Jalabert and J.-L. Pichard, J. Phys. I France {\bf 5}, 287 (1995).
	              
	              \bibitem{baranger-mello}
H. U. Baranger and P. A. Mello, Phys. Rev. Lett. {\bf 73}, 142 (1994).

\bibitem{jalabert}
R. A. Jalabert, J.-L Pichard, and C. W. J. Beenakker, Europhys. Lett. {\bf 27}, 255 (1994).

\bibitem{brouw} P. W. Brouwer, Phys. Rev. B {\bf 51}, 16878 (1995). 







\bibitem{edelmanrao} A. Edelman and N. Raj Rao, Acta Numerica {\bf 14}, 233 (2005). 













\end{thebibliography}
\end{document}